\documentstyle[epsf,floats,aps]{revtex}

\def\abs#1{\vert #1 \vert}
\newcommand{\beq}{\begin{equation}}
\newcommand{\eeq}[1]{\label{#1}\end{equation}}
\newcommand{\beqn}{\begin{eqnarray}}
\newcommand{\eeqn}[1]{\label{#1}\end{eqnarray}}
\def\nn{\nonumber}

\newcommand{\ba}{\begin{array}}
\newcommand{\ea}{\end{array}}

\newcommand{\Z}{{\cal Z}}

\newcommand{\Las}{{\cal L}}

\def\Las{{\cal L}}

\begin{document}


\title{On planar fermions with quartic interaction at finite temperature and density}

\author{D.C.\ Cabra$^{1,2}$ and G.L.\ Rossini$^1$}
\address{
$^{1}$Departamento de F\'{\i}sica, Universidad Nacional de la Plata,
C.C.\ 67, (1900) La Plata, Argentina.\\
$^{2}$ Facultad de Ingenier\'\i a, Universidad Nacional de Lomas
de Zamora,\\ Cno. de Cintura y Juan XXIII, (1832) Lomas de Zamora,
Argentina.}
\date{\today}
\maketitle
\begin{abstract}
\begin{center}
\parbox{14cm}{
We study the breaking of parity symmetry in the 2+1 Gross-Neveu
model at finite temperature with chemical potential $\mu$, in the
presence of an external magnetic field. We find that the
requirement of gauge invariance, which is considered mandatory in
the presence of gauge fields, breaks parity at any finite
temperature and provides for dynamical mass generation, preventing
symmetry restoration for any non-vanishing $\mu$. The dynamical
mass becomes negligibly small as temperature is raised. We comment
on the relevance of our observation for the gap generation of
nodal quasi-particles in the pseudo-gap phase of high $T_c$
superconductors.}
\end{center}
\end{abstract}


\section{Introduction}

The study of planar fermion systems has become an active area of
research in the last years, on its own right and because of the
many applications in strongly correlated electron systems as high
$T_c$ superconductors, quantum Hall effect systems, etc.
In particular, in the last few years several papers \cite{Gusynin} have analyzed
the role of quasi-particles in transport experiments in cuprates
in the under-doped regime in order to provide an explanation for
the unusual behaviour observed in \cite{Krishana}.
It was pointed there that under the influence of an external magnetic
field (transverse to the copper-oxide planes) the thermal
conductivity shows a plateau feature for magnetic fields larger
than a critical value. One possible scenario to explain  the
appearance of such a plateau is the opening of a gap in the
quasi-particle spectrum \cite{Gusynin,Liu}. More recent
experiments however seem to indicate the absence of the plateau in
the thermal conductivity for sufficiently clean samples
\cite{newexp}. In the light of this new data, it is the purpose of
the present paper to further discuss the issue of the opening of a
gap in the theory describing the so-called nodal quasi-particles,
in particular in the finite density case.

It is by now established that low temperature properties of
planar high $T_c$ superconductors are well described by a $d$-wave BCS theory
\cite{PLee}. Indeed, in a $d$-wave superconductor the energy gap (order
parameter) vanishes at certain points on the Fermi surface (nodes)
and hence low energy quasi-particle excitations are relevant for
low temperature transport measurements. Among the quasi-particle
excitations above the ground state we are interested in the
behaviour of the nodal quasi-particles. These are well defined
fermionic relativistic massless quasi-particles, which are assumed
to have quartic self interactions \cite{LeeWen}.

Our main point in the present paper is that in the presence of an external
magnetic field, fermionic fields have to be quantized in a
gauge-invariant framework and hence a parity anomaly could
naturally appear \cite{Redlich}. The parity breaking effect of course depends on the
external field configuration and is trivial (unobservable) in the case of
constant magnetic fields, unless a non vanishing chemical potential
($\mu$) is considered. Previous analysis \cite{Gusynin,Liu,Semenoff,Klimenko}
have either
considered the case with $\mu=0$ or have duplicated the number of
fermion components in such a way that the parity anomaly cancels
out. Our analysis, though motivated by high $T_c$ phenomenology,
is applicable to any planar fermion system.

We show that, at finite quasi-particle
density, the term that breaks parity in the effective action
changes dramatically the analysis of the gap equation. In
particular, the opening of a gap as a function of the external
magnetic field occurs at any finite temperature, thus providing
for dynamical mass generation and preventing symmetry restoration
for any non-vanishing $\mu$. It should be pointed that the effect
of this correction is of order $1/T$ and the gapless behaviour is
recovered at high temperatures.

\section{The model}

Let us consider
the following (Euclidean) $3$-dimensional fermionic Lagrangian with four
fermion interaction
\beq
\Las ={\bar \psi}^a ( \not \! \partial +
i e \not \!\! A + \gamma^0 \mu ) \psi^a +\frac{g}{2N}({\bar
\psi}^a\psi^a)^2,
\eeq{lag}
where $a=1, \cdots, N$ is a flavor index and the Fermi fields are
in an irreducible two component spinor representation. This model
field theory is known as the Gross-Neveu model. For the case of
interest in $d$-wave superconductors, nodal quasi-particles are
described by (\ref{lag}) with $N=4$, corresponding to the four
nodes of the energy gap. The gauge field $A_\nu$ $(\nu=0,1,2)$
represents an external background that for a constant transverse
magnetic field $B$ can be chosen as {\it e.g.} $A_0=0$, $A_i = -B
x_2 \delta_{i 1}$.

Apart from the usual gauge invariance
$A_\nu \rightarrow A^{(\lambda)}_\nu = A_\nu + e^{-1} \partial_\nu \lambda$,
$\psi \rightarrow \psi^{(\lambda)}=\exp(-i\lambda) \psi$,
the theory defined by eq.
(\ref{lag}) is at the classical level invariant under parity
transformations, which are defined as
\beqn
(x_0,x_1,x_2) & \rightarrow & (x_0,-x_1,x_2) \ ,
\nn\\
(A_0,A_1,A_2) & \rightarrow & (A_0,-A_1,A_2)
\nn\\
\psi & \rightarrow & \gamma_1 \psi \ .
\eeqn{parity}
Now, since in odd space-time dimensions the path integration measure cannot be
defined in a way that
preserves both gauge and parity invariance,
a parity violating contribution can arise if one adopts a
gauge invariant quantization. Indeed, from the definition of the
partition function $\Z[A_\nu]$, if
$\Z[A_\nu]=\Z[A^{(\lambda)}_\nu]$
one necessarily has to impose invariance of the fermionic measure
under gauge transformations $\psi  \rightarrow  \exp(-i\lambda) \psi$.
Due to the presence of an external magnetic field
we consider mandatory to quantize the theory in a gauge invariant
way, which then leads to the well known parity anomaly \cite{Redlich}.

It is also known that the parity anomaly can be overcome
by a slight change in the theory, consisting in the use of a
suitable reducible four component spinor representation for the
Fermi fields (as done in \cite{Gusy2}). As we said before, this change not
merely duplicates the number of components, but does also change
the interaction term \cite{Liu}.
We analyze in the following the case in which an irreducible spinor
representation and a gauge invariant regularization are chosen.

In order to study the opening of a gap in this system, which is
associated with the breaking of parity symmetry, we use the $1/N$
standard procedure to compute the effective potential for the
fermion system. One first introduces an auxiliary field $\sigma$
trading the quartic interaction term for a linear $\sigma$ vertex
\beq
\Las ={\bar \psi}^a ( \not \! \partial +
i e \not \!\! A + \gamma^0 \mu + \sigma) \psi^a  -\frac{N}{2g}
\sigma^2;
\eeq{HS}
the equation of motion for $\sigma$ sets the constraint
\beq
\sigma = \frac{g}{N}{\bar\psi}^a \psi^a.
\eeq{constraint}
Parity invariance of (\ref{HS}) at the classical level (or
alternatively consistency of eq.\ (\ref{constraint})) requires
that the field $\sigma$ changes as a pseudo-scalar under parity,
\beq
\sigma  \rightarrow  -\sigma \ .
\eeq{parity'}
The breaking of parity symmetry at the quantum level would be now
signaled by a non vanishing expectation value of the fermion
condensate. In order to search for this effect to leading order in
$1/N$ it is enough to consider constant values for $\sigma$.

The effective potential is defined as
\beqn
V^{eff}_{\beta ,\mu}[\sigma ] \equiv
- \frac{1}{\beta L^2} \log \left( \int {\cal
D}\bar \psi {\cal D} \psi \exp - \int_0^\beta d\tau \int d^2x
\left( {\bar \psi}^a ( \not \! \partial + i e \not \!\! A +
\gamma^0 \mu +\sigma) \psi^a -\frac{N}{2g}\ \sigma^2 \right)
\right)
\eeqn{defveff}
and the vacuum expectation for the fermion condensate can be found
from its minima, that is solving the gap equation
$\delta V^{eff}/\delta \sigma=0$.

We distinguish two different contributions to the effective
potential, one even in $\sigma$ defined as
\beq
V^{\rm even}[\sigma ]\equiv \frac{1}{2}(V^{eff}[\sigma ]+V^{eff}[-\sigma
])
\eeq{defVeven}
and the other odd in $\sigma$, which signals the breaking of
parity, defined as
\beq
V^{\rm odd}[\sigma ]\equiv \frac{1}{2}(V^{eff}[\sigma ]-V^{eff}[-\sigma
])
\eeq{defVodd}

The first contribution, $V^{even}$, can be computed by any method
that assumes that $V^{eff}$ depends on $\sigma^2$; in particular,
a detailed computation
was performed in \cite{Gusy2} using the
Schwinger proper time method. The renormalized result is
\beqn
V^{\rm even}_{\beta , \mu}[\sigma ]=\frac{N}{2\pi} \left[\frac{\Lambda}{2\sqrt{\pi}}(\frac{2\sqrt{\pi}}{g}-1)\sigma^2
-\frac{\sqrt{2}}{l^3} \zeta(-\frac{1}{2},\frac{(\sigma l)^2}{2}+1)
-\frac{\abs{\sigma}}{2l^2} \right]\nn\\
- \frac{N}{4 \pi \beta l^2}
\left\{ \rule{0cm}{.6cm}
\log\left( 1+\exp (-2 \beta\vert\sigma\vert) +2 \exp (-\beta\vert\sigma\vert) \cosh
(\beta\mu)\right)+
\right.
\nn\\
\left.
2\sum_{n=1}^{\infty}
\log\left( 1+\exp \left(-2 \beta\sqrt{\sigma^2+\frac{2n}{l^2}}\right) +
2 \exp \left(-\beta\sqrt{\sigma^2+\frac{2n}{l^2}}\right) \cosh
(\beta\mu)\right)
\right\},
\eeqn{Veven}
where $l=1/\sqrt{\vert e B\vert}$ and $\Lambda$ is an UV cutoff.


The parity violating contribution to the effective action for a
fermion system at finite temperature in a gauge background has
been recently computed in exact form for constant field strength
configurations in \cite{Rossini}. The result,
originally presented without consideration of a chemical
potential, can be straightforwardly applied to the case at hand,
since the chemical potential $\mu$ in eq.\ (\ref{defveff}) plays
the same role as an imaginary time component of the gauge field
and a constant $\sigma$ plays the role of a mass term. In fact,
replacing the time component $e A_0$ in \cite{Rossini} by $i\mu$
one gets
\beq
V^{\rm odd}_{\beta , \mu}[\sigma ]=- \frac{N}{2\pi \beta l^2}
\ {\rm arctanh} \left(
\tanh(\frac{\beta\sigma}{2})\tanh(\frac{\beta\mu}{2})
\right).
\eeq{Vodd}

The complete expression for the effective potential is of course the sum of both
contributions, $
V^{eff}_{ \beta , \mu}[\sigma ]=
V^{even}_{ \beta , \mu}[\sigma ]+V^{odd}_{ \beta , \mu}[\sigma ]$

It is now easy to see that term in (\ref{Vodd}) changes the mass
generation picture completely at low temperatures. This is because
it is smooth at the origin and odd in $\sigma$ and hence shifts the minimum of the
effective potential away from zero, leading to a mass gap.
A numerical analysis of the gap equation $\delta V^{eff}/\delta \sigma=0$
confirms that there is no minimum at $\sigma =0$,
except at very high temperatures, where $V^{odd}$ is
subdominant (of order $1/T$) with respect to the even terms.

In order to explore the meaning of $V_{eff}$ we show in the
following plots striking qualitative differences between the gauge
invariant  and the parity conserving effective actions.
In Fig.\ 1 we show the parity conserving effective potential for
fixed magnetic field and chemical potential for a range of
temperatures where the transition between massless and massive
regimes is apparent. In Fig.\ 2 we plot the gauge invariant
effective potential for the same range of parameters, in which
case the theory is always massive. In Fig.\ 3 we include higher
temperatures so as to show the tendency to symmetry restoration.
In all these figures we plot dimensionless quantities in terms of
an arbitrary mass scale. In particular we chose
$\Lambda=\sqrt\pi$, $g=0.9 \sqrt\pi$, $\mu=0.1$ and $eB=1$ \cite{Gusy2}.

\begin{figure}
\hbox{%
\epsfxsize=4.4in \hspace{3.5 cm} \epsffile{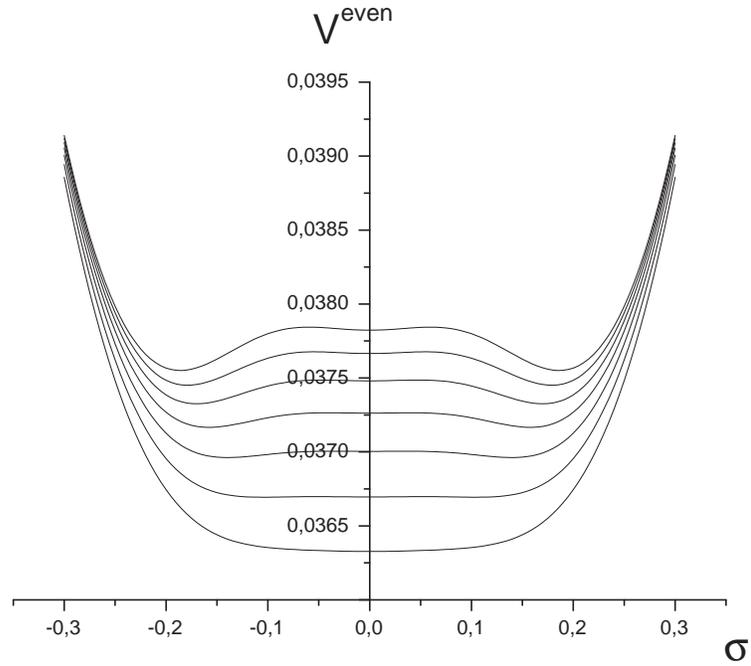}}
\vspace{0.3cm} \caption{Parity conserving effective potential,
showing the transition from massless to massive regime. Coupling
$g$, magnetic field and chemical potential are kept fixed. Higher
temperatures show symmetry restoration (lower curves).
\label{fig1} }
\end{figure}

\begin{figure}
\hbox{%
\epsfxsize=4.4in \hspace{3.5 cm} \epsffile{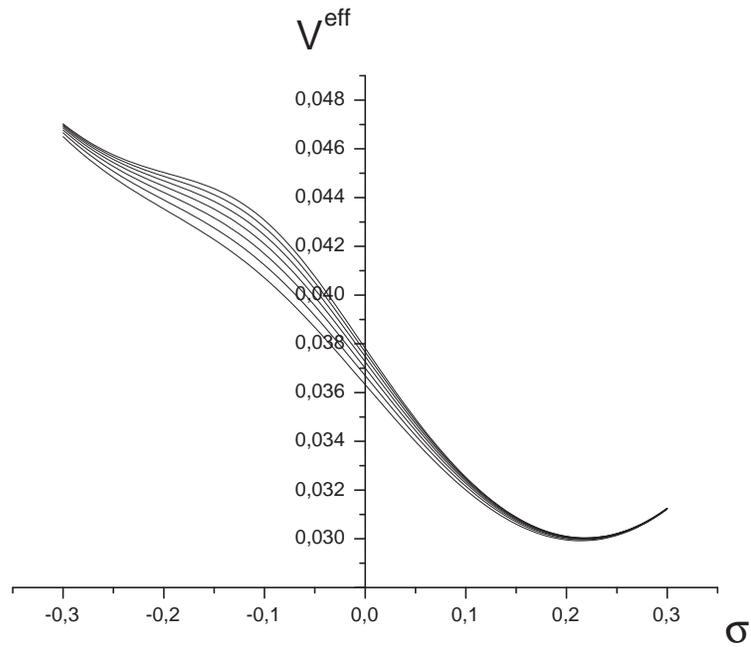}}
\vspace{0.3cm} \caption{Complete effective potential, showing
persistence of symmetry breaking for the same range parameters of
Fig.\ 1. Temperature grows from top to bottom. \label{fig2} }
\end{figure}

\begin{figure}
\hbox{%
\epsfxsize=4.4in \hspace{3.5 cm} \epsffile{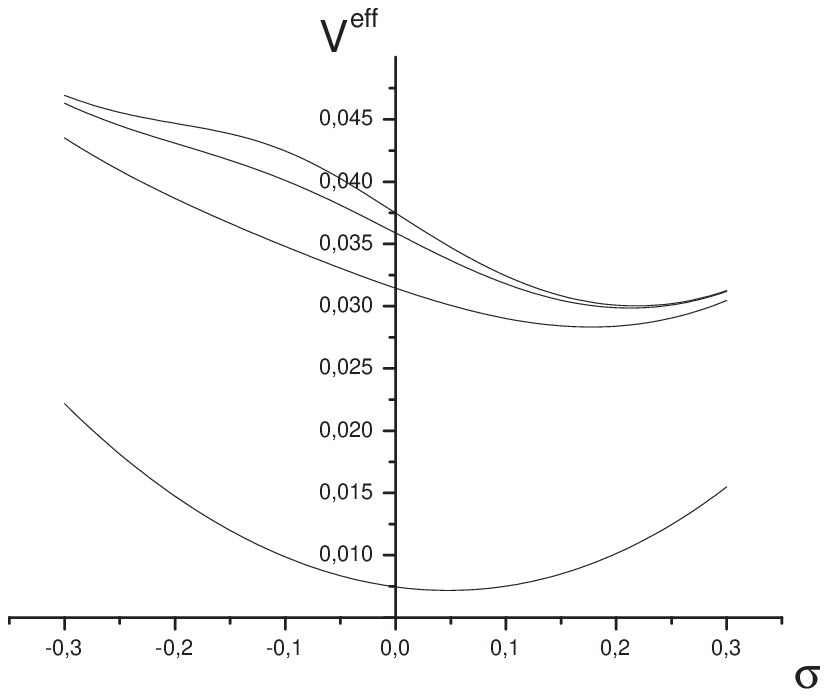}}
\vspace{0.3cm} \caption{Complete effective potential for a wider
range of temperatures, indicating the irrelevance of the parity
breaking contribution at high temperature (lower curve).
\label{fig3} }
\end{figure}

\section{Conclusions}

In the present paper we have analyzed the consequences of the parity anomaly on the
usual picture of dynamical mass generation (associated with parity symmetry
breaking) for planar fermions with quartic interactions at finite
temperature and density. Our main observation is that the
inclusion of a finite chemical potential prevents the appearance of a symmetric
phase for arbitrarily small magnetic fields.

One important context where our result could be of relevance is in
the interpretation of the plateau feature observed in thermal
conductivity curves in \cite{Krishana} and further discussed in
\cite{newexp} where contradicting evidence was reported. Within
our point of view for the description of nodal quasi-particles a
mass gap is present for any non-vanishing value of the external
magnetic field and chemical potential. Then, there could be no sharp changes in the
thermal conductivity due to nodal quasi-particle behaviour. This
result is consistent with the measures reported in \cite{newexp}
where no clear evidence for a plateau feature was observed. In the
case that further experiments confirm the appearance of such a
plateau in the thermal conductivity, our result would rule out the
mechanism proposed to explain this phenomenon as due to the
opening of a quasi-particle mass gap for a magnetic field above a
critical value.

\smallskip

We are grateful to C.D.\ Fosco, E.F.\ Moreno and F.A.\ Schaposnik
for useful discussions. The authors thank CONICET and Fundaci\'on
Antorchas (Grants No. A-13622/1-106 and 13887-73) for financial support.


\begin{references}

\bibitem{Gusynin} See {\it e.g.} E.J.\ Ferrer, V.P.\ Gusynin, V.\ de la Incera,
preprint hep-ph/0101308 and references therein.

\bibitem{Krishana} K.\ Krishana {\it et al},
Science {\bf 277}, 83 (1997).

\bibitem{Liu} W.V.\ Liu, Nucl.\ Phys.\ {\bf B 556}, 563 (1999).

\bibitem{newexp} Y.\ Ando {\it et al}, Phys.\ Rev.\ B {\bf 62}, 626 (2000).

\bibitem{PLee} P.A.\ Lee, preprint condmat/9812226.

\bibitem{LeeWen} P.A.\ Lee, X.-G.\ Wen,  Phys.\ Rev.\ Lett.\ {\bf 78}, 4111 (1997)

\bibitem{Redlich} N. Redlich, Phys.\ Rev.\ Lett.\ {\bf 52}, 18 (1984);
Phys.\ Rev.\ D {\bf 29}, 2366 (1984).

\bibitem{Semenoff} G.W.\ Semenoff, I.A.\ Shovkovy, L.C.R.\
Wijewardhana, Mod.\ Phys.\ Lett.\ {\bf A13}, 1143 (1998).

\bibitem{Klimenko}  V. Ch. Zhukovsky, K. G. Klimenko, V. V. Khudyakov, D.
Ebert, JETP Lett. 73 (2001) 121-125.

\bibitem{Gusy2} V.P.\ Gusynin, V.A.\ Miransky, I.A.\ Shovkovy,
Phys.\ Rev.\ Lett.\ {\bf 73}, 3499 (1994);
Phys.\ Rev.\ D {\bf 52}, 4718 (1995).

\bibitem{Rossini} C.\ Fosco, G.L.\ Rossini, F.A.\ Schaposnik,
Phys.\ Rev.\ Lett.\ {\bf 79}, 1980 (1997); {\bf 79}, 4296(E)
(1997); Phys.\ Rev.\ D {\bf 56}, 6547 (1997).

\end{references}
\end{document}